\RequirePackage{fix-cm}
\documentclass[smallextended]{svjour3}       
\smartqed  
\usepackage{graphicx}
 \usepackage{placeins}

\usepackage{tikzfig}

\usepackage{stmaryrd}

\usepackage{tkz-graph}
\GraphInit[vstyle = Shade]
\tikzset{
  LabelStyle/.style = { rectangle, rounded corners, draw,
                        minimum width = 2em, fill = yellow!50,
                        text = red, font = \bfseries },
  VertexStyle/.append style = { inner sep=5pt,
                                font = \Large\bfseries},
  EdgeStyle/.append style = {->, bend left} }

\usepackage{lineno,hyperref}
\usepackage{algorithm}
\usepackage[noend]{algorithmic}
\usepackage{amssymb}
\usepackage{graphicx}
\usepackage{amsmath}
\usepackage{algorithm}
\usepackage[noend]{algorithmic}
\usepackage{url}
\usepackage{caption}
\usepackage{multirow}
\usepackage{threeparttable}
\usepackage{color}
\usepackage{lineno}

\usepackage{tkz-graph}

\tikzset{
  LabelStyle/.style = { rectangle, rounded corners, draw,
                        minimum width = 2em, fill = yellow!50,
                        text = red, font = \bfseries },
  VertexStyle/.append style = { inner sep=5pt,
                                font = \Large\bfseries},
  EdgeStyle/.append style = {->, bend left} }
  
  \usetikzlibrary{shapes.geometric}
\usetikzlibrary{positioning}

\usetikzlibrary{matrix,calc,shapes}
\tikzset{
  treenode/.style = {shape=rectangle, rounded corners,
                     draw, anchor=center,
                     text width=3em, align=center,
                     top color=white, bottom color=blue!20,
                     inner sep=1ex},
  voters/.style = {treenode, diamond, inner sep=0pt},
  miners/.style      = {treenode}
}
\newcommand{\commit}{edge node [above] {commit}}
\newcommand{\commitOpen}{edge node [left] {commit-open}}

\newcommand{\verifyA}{edge node [above] {verify}}
\newcommand{\verifyB}{edge node [below] {verify}}

\begin{document}

\title{Bit Commitment for Lottery and Auction on Quantum Blockchain
}


\author{Xin Sun        \and
        Piotr Kulicki \and Mirek Sopek 
}


\institute{Xin Sun \at
              Department of the Foundations of Computer Science, 
              the John Paul II Catholic University of Lublin, Poland\\
              \email{xin.sun.logic@gmail.com}           
           \and
           Piotr Kulicki \at
            Department of the Foundations of Computer Science, 
            the John Paul II   Catholic University of Lublin, Poland \\
              \email{kulicki@kul.pl}          
              \and 
              Mirek Sopek \at
            Makolab SA, Lodz, Poland \\
           \email{sopek@makolab.com}
}

\date{Received: date / Accepted: date}

\maketitle

\begin{abstract}

This paper propose a protocol for lottery and a protocol for auction on quantum Blockchain. Our protocol of lottery satisfies randomness, unpredictability, unforgeability, verifiability, decentralization and  unconditional security. Our protocol of auction satisfies bid privacy,
posterior privacy, bids’ binding, decentralization and unconditional security. Except quantum Block-chain, the main technique involved in both protocols is quantum bit commitment.   
Since both quantum blockchain and quantum bit commitment can be realized by the current technology, our protocols are practically feasible.

\end{abstract}

\section{Introduction}

A blockchain is a distributed, transparent and append-only ledger of cryptographically linked units of data (blocks), which incorporates mechanisms for achieving consensus over the blocks of data in a large decentralised network of nodes which do not trust each other. It is a ledger in the sense that the data entries stored on the blockchain can be considered as generalized transactions.  It is a distributed system in the sense that all miners (the nodes that are in charge of updating the ledger) have separated, identical copies of the ledger. One of the most prominent applications of Blockchain technology is to enable the creation and distribution of cryptocurrencies, such as Bitcoin \cite{Nakamoto08}. Another important application is the implementation of smart contracts \cite{Szabo97}, which are enforceable, irrefutable agreements among mutually distrusting peers, which do not imply a trusted third party for their affirmation and administration mechanism.  

The power of quantum computers and the capabilities of existing quantum algorithms \cite{Shor97} represent a threat to most of the existing public-key cryptographic systems. The current predictions \cite{Mosca18} assume that by 2026 the chance of the practical availability of quantum computers is about 15\% and by 2031 the chance grows to 50\%. As almost all existing Blockchain implementations have very deep reliance on the public-key digital signatures and as they are used for the transfer of value, they are particularly vulnerable to the attack of quantum computers. As pointed out by Fedorov \textit{et al.} \cite{Fedorov18Nature}, Blockchain technology as we know it today may founder unless it integrates quantum technologies.   

There is a considerable amount of research related to the quantum-safe Blockchain \cite{Kiktenko17,Aggarwal18,Stewart18,Sun19blockchain,Sun19vote} which could withstand attacks powered by forthcoming quantum computers. One of the most prominent proposals is the Quantum-secured Blockchain (QB) developed by Kiktenko \textit{et al.} \cite{Kiktenko17}.
Due to the application of unconditionally secure message authentication based on quantum key distribution methodology, QB is immune to the attacks of quantum computers.  The major limitation of QB is that the consensus protocol it adopts is not efficient, because 
 it becomes exponentially data-intensive if a large number of cheating miners is present.  
This limitation is overcome in  \cite{Sun19blockchain}, where 
a new consensus protocol is reported, exhibiting only quadratic dependence of resources on the number of miners. 
In \cite{Sun19vote} the usage of quantum blockchain is illustrated by designing a simple voting protocol based on it. To further demonstrate the power and application potential of quantum blockchain, in this paper we  present protocols for lottery and auction based on it.

Lottery is a multi-billion dollar industry \cite{Isidore15}. In general, in a lottery there is an authority and a number of players. Players buy tickets to participate in the game. Then a random process is used to determine the winning tickets. In many lotteries, the revenue is huge and so is the incentive to cheat. In order to ensure fair play and the trust of players, an ideal lottery  protocol \cite{Chow05,Bentov14,Andrychowicz14,Bartoletti17,Grumbach17,Miller17} should satisfy the following requirements:

\begin{enumerate}

\item  Randomness. All tickets are equally likely to win. 

\item Unpredictability. No player can predict the winning ticket.

\item  Unforgeability. Tickets cannot be forged. Especially, it is impossible to create a winning ticket after the outcome of the random process is known.

\item  Verifiablity. The number  and the revenue of winning tickets are publicly verifiable.

\item Decentralization. The random process does not rely on a single authority.

\end{enumerate}

Lottery protocols that satisfy the above requirements already exist \cite{Chow05,Grumbach17}. 
With the advent of the quantum computing technology, it is reasonable to further require lottery protocols to satisfy:

\begin{enumerate}

\item[6.] Unconditional security. Even an adversary with unlimited power of computation cannot rig the lottery.

\end{enumerate}

\noindent
Although quantum coin flipping \cite{Goldenberg99,Spekkens02,Nayak03,Ambainis04,Nguyen08,Silman11,Hanggi11,Nayak16}, a specific form of lottery, has been researched in the past 20 years, only randomness and the unconditional security have been studied in those works, while other properties of lottery have rarely been addressed in the context of quantum coin flipping.
In this paper, for the first time, we design a lottery protocol which satisfies all the above requirements.

Auction is an even more important business 
in the sense that trillions of dollars are transferred by auctions. 
An auction is a process of buying and selling goods by offering them up for bid, taking bids, and then selling the item to the buyer who offers the highest bid. 
In general, there are two types of auction: sealed-bid auction and non-sealed-bid auction. 
The main advantage of the sealed-bid auction lies in the fact that
no buyer gets to know the bids offered by other buyers.
In the literature \cite{Brandt03,Brandt06,Montenegro13} it is acknowledged that an ideal sealed-bid auction must satisfy the following properties:

\begin{enumerate}

\item Bid privacy. The submitted bids are not visible to other buyers during the bidding phase.

\item Posterior privacy. The losing bids are not revealed to the public. In other words, only the seller knows all losing bids and their corresponding buyers.

\item Bids’ binding. Buyers cannot deny or change their bids once they are committed.

\end{enumerate}

In the setting of quantum blockchain, it is reasonable to require that the auction protocol further satisfies the following properties:

\begin{enumerate}

\item[4.] Decentralization. The process of auction does not rely on a single trusted third party.

\item[5.] Unconditional security. Even an adversary with unlimited power of computation cannot manipulate the process of auction.

\end{enumerate}

While blockchain-based auction \cite{Blass18,Galal18} does satisfy decentralization and quantum auction  \cite{Liu16,Zhang18}
does satisfy unconditional security, no existing auction protocol satisfies both of these properties.
The auction protocol we are going to propose satisfies all the above properties.

Except quantum blockchain, the main technique that we will use is quantum bit commitment.
We first review some background knowledge of quantum blockchain and  quantum bit commitment in Section \ref{Background}. We then present our lottery protocol in Section \ref{Lottery on quantum blockchain} and auction protocol in Section \ref{Auction}.
We finish this paper in Section \ref{Conclusion and future work} with conclusions and remarks on the future work. 

\section{Background}\label{Background}

\subsection{Quantum Blockchain}

The concept of quantum blockchain presented in \cite{Kiktenko17,Sun19blockchain,Sun19vote}, which we are going to explore for our lottery and auction protocols, assumes that each pair of nodes is connected by a quantum channel and a classical channel.
Every pair of nodes can establish a sequence of secret keys by using the quantum key distribution \cite{Bennett84} mechanisms. Those keys will later be used for secure communication.

%

Updates (new transactions or new messages) on blockchain are initiated by those nodes that wish to append some new data to the chain.  Each miner checks the consistency of the update with respect to their local copy of the database and works out a judgment regarding the update's admissibility. 
Then all the miners apply a consensus algorithm to the update, arriving at a consensus regarding the correct version of the update.

In this paper, we will consider quantum blockchain on a high level, omitting its detailed structure and mechanism, and taking advantage of its following desired properties:

\begin{enumerate}
\item Every node is a (small scale) quantum computer which can run some quantum computation on a small number of qubits. More specifically, nodes are capable of performing the quantum computation involved in at least one quantum bit commitment protocol.

\item The communication between different nodes is unconditionally  secure.

\item There is a consensus algorithm  which can be used by all miners to achieve consensus. The consensus mechanism is immune to attacks. A general definition of the consensus algorithm is given as the following.

\end{enumerate}

\begin{definition} [consensus algorithm] An algorithm among $n$ parties, in which every party $p$ holds an  input value
$x_p \in D$ (for some finite domain $D$)  and
eventually decide on an output value in $y_p\in D$, is said to achieve consensus if the algorithm guarantees that the output value of all honest parties are the same.


\end{definition}

\subsection{Quantum Bit Commitment}

Bit commitment typically consists of two phases, namely: commitment and opening.
In the commitment phase, Alice, the sender, chooses a bit $a$ ($a = 0$ or $1$) which she wishes to commit to Bob, the receiver. Then Alice presents Bob some evidence about the bit. The committed bit cannot be known to Bob prior to the opening phase. Later, in the opening phase, Alice discloses 
some information needed for the reconstruction of $a$. Then Bob reconstructs a bit $a'$ using Alice's evidence and the disclosure. A correct bit commitment protocol will ensure that $a' = a$. A bit commitment protocol is concealing if Bob cannot know the bit Alice committed before the opening phase, and is binding if Alice cannot change the bit she committed after the commitment phase. 

The first quantum bit commitment (QBC) protocol was proposed in 1984 by Bennett and Brassard  \cite{Bennett84}. 
 A number of QBC protocols have been designed to achieve unconditional security, such as those of \cite{Brassard90,Brassard93}.
Although according to the Mayers-Lo-Chau (MLC) no-go theorem \cite{Mayers97,LoChau97,Sun20Axioms}, unconditionally secure QBC cannot be achieved within the theory of quantum mechanics, scientists have found ways to avoid this negative result in the past two decades. For example, cheat-sensitive quantum bit commitment (CSQBC) protocols \cite{Hardy04,Buhrman08,Shimizu11,Li14,Zhou19} and relativistic QBC protocols \cite{Kent11,Kent12,Lunghi13,Adlam15,Lunghi15,Verbanis15} have been developed. With well-designed mechanisms of punishment, the CSQBC protocols can be useful in practice and resilient to the attack of quantum computers.  
Relativistic QBC protocols achieve unconditional security by making use of the power of relativity theory. In \cite{Verbanis15}, the authors implemented a relativistic QBC protocol in which the bit is concealed for 24 hours. 
Another practically useful QBC can be found in He \cite{He11,He14}, who proposed a QBC protocol based on the use of Mach-Zehnder interferometer. 
He's protocol is also implementable by the current technology.
To sum up, practically useful QBC protocols are already available and are ready for  applications to other computational tasks.

The following is an abstract yet rigorous definition of QBC, which can be found in Sun \textit{et al.} \cite{Sun20Axioms} and will be used in this paper.

\begin{definition}[quantum bit commitment] \label{QBCDef}  
A quantum bit commitment protocol consists of the following:

\begin{enumerate}

\item[(1)] Two finite dimensional Hilbert spaces $A$ and $B$.

\item[(2)] A function $commit: \{0,1\} \mapsto A \otimes B$.

\item[(3)] Two pure states $|c_0\rangle, |c_1\rangle \in  A \otimes B$, in which $|c_i \rangle =commit(i)$ is the commitment of $i$.

\item[(4)] A quantum operation (\textit{i.e.} completely positive, trace-preserving super operator) $Open$ on $A \otimes B$ such that $Open(|c_0\rangle \langle c_0|) \neq Open(|c_1\rangle \langle c_1|) $.


\end{enumerate}
This QBC protocol is 
\textit{concealing} if  $Tr_A (|c_0\rangle \langle c_0|) = Tr_A (|c_1\rangle \langle c_1|)$. It is \textit{binding} if there is no unitary $U$ on $A$ such that $(U \otimes I_B )|c_0\rangle = |c_1\rangle$.

\end{definition}

\section{Lottery on quantum blockchain}\label{Lottery on quantum blockchain}

\begin{figure}

\begin{center}

\begin{tikzpicture}[-latex]
  \matrix (chart)
    [
      matrix of nodes,
      column sep      = 5em,
      row sep         = 5ex,
      column 1/.style = {nodes={voters}},
      column 2/.style = {nodes={miners}}
    ]
    {$player_1$          &  $miner_1$       \\
      $player_2$       & $miner_2$         \\
    };

   \draw (chart-1-1) \commit (chart-1-2);
   \draw (chart-2-1) \commit (chart-2-2);

   \draw (chart-2-1) \commit (chart-1-2);
   \draw (chart-1-1) edge (chart-2-2);
   
  \path [draw, -] (chart-1-2) -- node [right] {CA} 
        (chart-2-2);

\end{tikzpicture}

\caption{A network of players and miners:  Players commit their tickets to miners. Miners use a consensus algorithm (CA) to achieve consensus about the players' tickets.}
\label{A network of player and miners}

\end{center}

\end{figure}
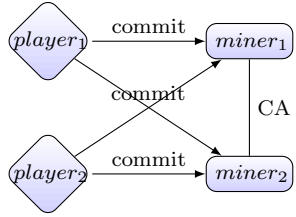

Now let us present our lottery protocol.
In the setting of lottery, we assume there are $n$ players and every ticket of the lottery is an $m$-bit string. Our lottery protocol 
consists of 3 phases: the ticket purchasing
phase, the ticket agreement phase and the winner determination phase. Figure \ref{A network of player and miners} presents simplified visualization of our protocol.

\begin{enumerate}

\item Ticket purchasing:

\begin{enumerate}
\item  For every player $p_i \in\{p_1,\ldots ,p_n \}$, to purchase a ticket $T_i$,  $p_i$ uses QBC to commit $T_i$ to all miners. At the end of this phase, every miner possesses a list of commitments $(commit(T_1), \ldots, commit(T_n) )$.

\end{enumerate}

\item Ticket agreement: 

\begin{enumerate}

\item Every player opens his commitment to every miner, so that the commitments in every miner's possession change to $(Open(commit(T_1)), \ldots,$ $ Open(commit(T_n))) $, which essentially equals to $(T_1, \ldots, T_n)$.

\item All the miners run a consensus algorithm to achieve a consensus on the tickets $(T_1, \ldots, T_n)$ purchased by players.
Every miner adds $(T_1, \ldots, T_n)$ to his local copy of the blockchain. 

\end{enumerate}

\item  Winner determination:

\begin{enumerate}

\item The winning ticket is calculated by bit-wise XOR: $T=T_1 \oplus \ldots  \oplus T_n$.

\item A player's revenue is determined by the Hamming distance between his ticket and the winning ticket $T$. The closer  his ticket is to the winning ticket, the higher is his revenue.\footnote{A specific rule of revenue which satisfies this principle is beyond the scope of this paper and is left for future work.} 

\end{enumerate}

\end{enumerate}

\subsection{Analysis}

Our lottery protocol satisfies the following  requirements: 

\begin{enumerate}

\item  \textbf{Randomness}. 

The winning ticket is calculated by bit-wise XOR. For every index $j\in \{1,\ldots,m\}$ in the winning ticket, $T[j]=1$  iff $T_1[j] \oplus \ldots \oplus T_n[j] =1$. Therefore, the probability of  $T[j]=1$ is the same as $T[j]=0$.

\item \textbf{Unpredictability}. 

To predict the winning ticket a player has to know all tickets before they are opened.
The concealing property of QBC ensures that even miners cannot know the players' tickets before they are opened. Since tickets are only sent to the miners by QBC, the probability that a player knows all tickets is even lower than the probability that a miner knows them.

\item  \textbf{Unforgeability}. 

The binding property of QBC ensures that it is impossible to change a ticket after the ticket purchasing phase.
 

\item  \textbf{Verifiablity}. 

This is because the quantum blockchain is a transparent database. After the ticket agreement phase the list $(T_1,\ldots T_n)$ is added to the blockchain. Every player can read all the other players' tickets and calculate the winning ticket by himself.

\item \textbf{Decentralization}. 

The random process does not rely on a single authority. 
Every player's ticket essentially affects the calculation of the winning ticket.
Moreover, the calculation of the winning ticket does not rely on a single miner, but on all miners.

\item  \textbf{Unconditional security}. 

Even an adversary with an unlimited power of computation cannot manipulate the lottery protocol. The concealing and binding property of QBC does not rely on any computational assumption. Nor does the security of the consensus algorithm.
The unconditional security of 
the ledger is further guarantied by the unconditional security of the digital signature schemes adopted by 
quantum Blockchain.

\end{enumerate}

\section{Auction on quantum blockchain}\label{Auction}

In our protocol of auction, we assume three types of participants: one seller $S$, $m$ buyers $\{B_1,\ldots,B_m\}$ and $n$ miners $\{M_1,\ldots,M_n\}$. Our protocol works as follows: First all buyers send their bids to the seller. Then the seller calculates which buyer is the winner. Finally, all miners verify the seller's calculation. Figure \ref{A network of the seller, buyers and miners} is a brief visualization of the process of auction. There are 5 phases in our protocol.

\begin{figure}

\begin{center}

\begin{tikzpicture}[-latex]
  \matrix (chart)
    [
      matrix of nodes,
      column sep      = 5em,
      row sep         = 5ex,
       column 1/.style = {nodes={voters}},
      column 2/.style = {nodes={voters}},
      column 3/.style = {nodes={miners}}
    ]
    {  & $buyer_1$          &  $miner_1$       \\
    $seller$ &  &  \\
      & $buyer_2$       & $miner_2$         \\
    };

   \draw (chart-1-2) \commit (chart-1-3);
    \draw (chart-1-2) \commitOpen (chart-2-1);
   \draw (chart-3-2) \commit (chart-3-3);
   \draw (chart-3-2) \commitOpen (chart-2-1);

\draw (chart-2-1) \verifyB (chart-1-3);
 \draw (chart-2-1) \verifyA (chart-3-3);       
   \draw (chart-3-2) \commit (chart-1-3);
   \draw (chart-1-2) edge (chart-3-3);
   
  \path [draw, -] (chart-1-3) -- node [right] {CA} 
        (chart-3-3);

\end{tikzpicture}

\caption{A network of the seller, buyers and miners.}
\label{A network of the seller, buyers and miners}

\end{center}

\end{figure}
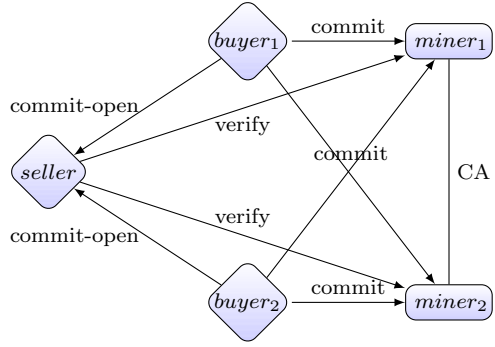

\begin{enumerate}
\item The bidding phase: Every buyer $B_i$ commits his bid $b_i$ to the seller and to all miners $M_j$, where $b_i$ is a positive integer.

\item  The opening phase: Every buyer opens his bid to the seller. 

\item Decision phase: The seller calculates the winning bid, which is the highest bid (if there is a tie, 
then one of the maximal bids is chosen randomly), and the winning buyer, who has offered the winning bid. 

\item Verification phase: In this phase  the seller $S$ and every miner $M_j$ run the following procedure to convince $M_j$ that $S$ has chosen the valid winner:

\begin{enumerate}

\item $S$ sends the information about the winning buyer $B_w$ and his bid $b_w$ to the miner $M_j$.

\item $S$ permutes losing bids to obtain a new list of $m-1$ bids $(b'_1,\ldots, b'_{m-1})$.

\item $S$ sends $b'_1,\ldots, b'_{m-1}$ to $M_j$.

\item $M_j$ first checks if $b_w\geq b'_k$ for all $k\in \{1,\ldots, m-1\}$. If yes, then $M_j$ sends $(b_w,  b'_1,\ldots, b'_{m-1})$ to all buyers. Otherwise, $M_j$ sets $S$ as a cheater and outputs $\bot$.

\item After receiving $(b_w,  b'_1,\ldots, b'_{m-1})$, every buyer $B_i$ checks if his bid is in the list, \textit{i.e.} there is some $b'_k = b_i$.
If yes, then $B_i$ sends the message ``valid'' to $M_j$. Otherwise $B_i$ opens $b_i$ to $M_j$. $M_j$ then sets $S$ as a cheater and outputs $\bot$.

\item If $M_j$ does not output $\bot$, then the seller passes the verification phase. The output of $M_j$ is now $(b_w,  b'_1,\ldots, b'_{m-1}, B_w)$

\end{enumerate}

\item Publication phase: 
All miners run the consensus algorithm to achieve consensus on the output of the verification phase. The consensus  is then added to the blockchain.

\end{enumerate}

\subsection{Analysis}

Our auction protocol satisfies the following  requirements: 

\begin{enumerate}

\item[1.] \textbf{Bid privacy}. 

Every buyer only commits and opens his bids to the seller. Therefore, no buyer knows other buyers's bid.

\item[2.] \textbf{Posterior privacy}. 

What is added to the blockchain is the winning buyer and his bid, as well as  a permuted list of losing bids. Therefore, no losing buyer's bid is revealed.

\item[3.] \textbf{Bids’ binding}. 

Binding property of quantum bit commitment ensures that buyers cannot deny or change their bids once they are committed.

\item[4.] \textbf{Decentralization}. 

There are in total $n$ miners. The process of auction does not rely on a single miner. 

\item[5.] \textbf{Unconditional security}. 


As in the case of our lottery protocol, even an adversary with an unlimited power of computation cannot manipulate the auction protocol because the security of the quantum bit commitment and consensus algorithm does not depend on computational complexity.
The unconditional security of the ledger relies on quantum Blockchain properties.

\end{enumerate}

\section{Conclusions and future work}
\label{Conclusion and future work}

This paper proposes a lottery protocol and an auction protocol based on quantum bit commitment and quantum blockchain.
 These  protocols satisfy all the important properties of distributed lottery/auction and are implementable by the current technology.
 

In the future, we are interested in applying quantum blockchain to the general field of multi-party  computation. We believe that quantum blockchain will provide new insights into these interesting tasks. We estimate that in the future more complicated protocols (smart contracts) on quantum blockchain will be designed. Developing a formal tool for the specification and verification of smart contracts on quantum blockchain is on our agenda. The recently developed categorical logic of quantum programs \cite{Sun20Entropy} seems to be a good starting point.


\begin{acknowledgements}
The project is funded by the Minister of Science and Higher Education within the program under the name ``Regional Initiative of Excellence'' in 2019-2022, project number: 028/RID/2018/19, the amount of funding: 11 742 500 PLN.
\end{acknowledgements}



\bibliographystyle{spmpsci}
\bibliography{biblio}

\end{document}